\documentclass[a4paper,twoside,reqno]{bjp}
\usepackage{graphicx}
\usepackage{cite}
\usepackage{amssymb,amsmath,amscd,amsthm}
\usepackage{times}
\usepackage{csquotes}

\usepackage[bookmarks=false]{hyperref}
\hypersetup{%
    colorlinks=true,        
    linkcolor=blue,          
    citecolor=blue,         
    urlcolor=blue           
    }

\usepackage{geometry}
\allowdisplaybreaks

\begin{document}

\title{Study of fusion reactions of light nuclei at low energies using complex nucleon-nucleus potential function}

\runningheads{Md. A. Khan et al. }{Study of fusion reactions of light nuclei at low energies .....}

\begin{start}{%
\author{Md. A. Khan$^*$}{1},
\author{S. H. Mondal}{1},
\author{M. Alam}{1},
\author{M. Hasan}{1}

\address{Dept. Of. Physics, Aliah University IIA/27 Newtown  Kolkata-700160 India}{1}

\received{Day Month Year (Insert date of submission)}
}

\begin{Abstract}
Nuclear fusion reactions, at energies, far below the Coulomb barrier play a significant role in the synthesis of light elements in the primordial nucleosynthesis as well as in the interior of compact stellar objects. Many different kinds of nuclear reactions are occurring simultaneously inside the stellar core depending upon the density and temperature conditions of the nuclear plasma along with other relevant parameters of these stars. Nuclear fusion reactions in the energy range ($E\sim$ 1 eV to few keV) can be explained successfully by quantum mechanical tunneling through
the mutual Coulomb barrier of interacting nuclei. The measurement of the cross-sections at extremely low energy is quite difficult because of the larger width of the Coulomb barrier, which results in a very small value of the reaction cross-section. Hence, any improvement in the data on astrophysical S-factors for the light nuclei fusion may give a better picture of the elemental abundance in nucleosynthesis. In this work, we have theoretically investigated the energy dependence of fusion cross-sections and astrophysical S-factors for fusion reaction of light nuclei like D-D and p-$^{11}$B using complex Gaussian nuclear potential with adjustable depth and range parameters plus the mutual Coulomb interaction of the interacting nuclei. Numerical computation of the observables is done in the framework of the selective resonant tunneling model approach. The results of our calculation are compared with those found in the literature. 

\end{Abstract}

\begin{KEY}
Cross-section, Complex potential, SRTM, S-factor
\end{KEY}
\end{start}


\section{Introduction}
 Nuclear physics plays a vital role in many important phenomena occurring in the core to the crust of astrophysical objects, this leads to the development of this ever-hot discipline of physics called nuclear astrophysics. It was Bethe who in 1939 first worked on to describe various cycles of stellar evolution with a prime focus on the mechanisms of production of light elements like $^{2}$H, $^{3}$He, $\alpha$, etc in the Big-Bang nucleosynthesis \cite{bethe-1939}. Works of Bethe was later advanced by Burbidge 1957 \cite{burbidge-1957}. Primordial nucleosynthesis is found to be limited up to the production of the isotopes of iron (Fe). Elements heavier than iron are found to be formed through various other processes like neutron capture- Reifarth et al. 2018 \cite{reifarth-2018}, neutrino-induced reactions- Alvarez-Ruso et al.  2018 \cite{alvarezruso-2018}, explosive events in supernovae- Wiescher et al. 2012 \cite{wiescher-2012}, and the rapid-neutron process in neutron-star mergers- Thielemann et al. 2017 \cite{thielemann-2017}. A systematic review of the most critical nuclear reactions under several nucleosynthesis environments and the status of those reactions giving insight into the speciﬁc uncertainties associated with their reaction rates has been presented by Wiescher et al. 2012\cite{wiescher-2012}. The review reports a large quantity of data on nuclear reactions involving chargeless particles (neutrons) and charged particles (protons and alphas) relevant to nucleosynthesis networks. Processes that play a major role in nuclear reactions involving charged particles are the transfer and capture processes. The transfer process is solely controlled by the strong nuclear interaction while electromagnetic interaction takes the credit for the capture process. When both channels are open the capture cross-section is always smaller than the transfer cross-section. A standard theoretical model for the evaluation of the reaction rates involves energy-dependent reaction cross-section ($\sigma(E)$) as a part of the integral expression and one of the major challenges nuclear physicists ever faces is the evaluation of the cross-section at stellar energies which are usually much smaller than the height of the mutual Coulomb barrier of the interacting nuclei. As the cross-sections are too small to be determined in the laboratory, direct measurement in the low energy regime ($E\sim$ 1eV to few keV) is not possible. And theoretical evaluation of astrophysical S-function ($S(E)$) is also model-dependent, hence the uncertainties in the computed S-factor can be significant \cite{yakovlev-2010} in nuclear physics. As discussed by Broggini et al. 2010 \cite{broggini-2010}, there exist few experimental techniques developed during the last few decades, and a sound theoretical justification is often needed for a successful explanation of the experimental findings at the stellar energies. Several indirect techniques have also been developed during the last couple of decades like the Trojan Horse method of Baur et al. 1986 \cite{baur-1986}, Tumino et al 2013 \cite{tumino-2013}, Spitalery et al. 2019 \cite{spitaleri-2019}, the Coulomb breakup method of Baur et al. 1986 \cite{baur-1986}, and the Asymptotic Normalization Coefficient method of Mukhamedzhanov et al. 2001 \cite{mukhamedzhanov-2001}. However, the methods referred above require a standard theoretical scheme to estimate the reaction cross-section from the captured data.
 
In the present work, we studied the energy dependence of nuclear fusion cross-section and astrophysical S-function for the fusion reaction of few light nuclei following an elegant theoretical model. A complex nuclear potential function is considered which facilitates the description of absorption inside the nuclear potential well. Here we adopt the selective resonant tunneling model (SRTM) of Li et al. 2000 \cite{li-2000} instead of the conventional compound nucleus model. Li et al was the first group to propose the SRTM model to compute data on fusion cross-section of D+T reaction using complex square-well and to compare their findings with the available experimental data.  Later in 2002, Li showed that this model also works well for D+T reaction up to 100 KeV \cite{li-2002}. Li et al again in 2004 \cite{li-2004} used the SRTM model to compute fusion cross-sections for D+D and D+$^{3}$He fusion reactions. Recently, Singh et al. 2019 \cite{sing-2019} applied this model considering a complex square-well nuclear potential to study fusion cross-section and astrophysical S-function for some light nuclear reactions like D+D, D+T, D+ $^{3}$He, p+$^6$Li, p+$^7$Li etc. 

In this work, we will use the complex Gaussian type potential function instead of the complex square-well potential as used by Li et al \cite{li-2000} to investigate some light nuclear fusion reactions like the D+D and p+$^{11}$B fusion at the deep sub-barrier energy region. The data obtained are to be checked with the well-known three-parameter and five-parameter fitting formula's from the NRL Plasma Formulary \cite{huba-2013} for reactions having fusion cross-section data available in the literature. The improved values of the astrophysical S-factor for light nuclei is expected to give a better and more clear picture of the elemental abundance in nucleosynthesis.

In section 2, we briefly discuss the theoretical method involved in the selective resonant tunneling model and its application to a light nuclear fusion reaction. Results and discussions will be presented in Section 3, and finally, we will present the summary and conclusions in section 4.
 
\section[]{Theoretical method}
The nuclear fusion process is supposed to occur via two independent processes: the first involves the penetration of the incoming projectile through the Coulomb barrier and the second involves the actualization of fusion reaction as described by Clayton et al. 1990 \cite{clayton-1990}. Resonant tunneling in the light nuclear fusion reaction is sequentially followed by tunneling and decay. A theoretical model based on the assumption of the fact that \enquote{decay is independent of tunneling} does not give a clear picture of the fusion process pointed out by Gamow in 1938 \cite{gamow-1938}. Selective resonance tunneling model (SRTM) is different from the well-known compound nucleus model (CNM) because, in the former one, the penetrating particle may still remember its phase unlike in the latter one in which the penetrating particle loses memory of its formation as described in the volume by Feshbach 1992 \cite{feshbach-1992}. The uniqueness of this model comes from the keyword 'selectivity'. This can be understood if we visualize absorption to act like damping in a resonance. The energy absorbed by a damping process is proportional to the product of the damping coefficient and the square of the amplitude of the oscillation. For a vanishing damping coefficient, the energy absorbed by the damping process is zero even if the resonance develops fully. On the other hand, for a very large value of damping coefficient, the damping process will destroy the resonance before it is fully developed. Thus, the energy absorbed by the damping process is still very small. Hence,
there must be some specific damping that makes the absorbed energy maximized. In a similar understanding, the fusion cross-section is proportional to the product of the depth of the imaginary component of nuclear potential and the square of the amplitude of the wave function inside the nuclear well \cite{frobrich-1996}; thus, there should be suitable damping represented by the depth parameter of the imaginary component of the nuclear potential to make the fusion cross-section maximized. We may call this suitable damping as the matching damping \cite{li-2000}. Inside the nuclear well, amplitude of deuteron wave function happens to be small in majority of the cases due to attenuation of the incoming wave by the Coulomb barrier. However there are fair chances that this attenuated wave fuction while making bouncing to and fro motion between the waalls of the nuclear well may interfere constructively with the incoming wave to build up sufficient amplitude. And when the amplitude of the enhanced wave function becomes comparable to the amplitude of the wave function outside the Coulomb barrier, resonace tunneling may occur. Hence, the constructive interference between the reflected and incoming wave function need a particular energy level (or frequency) and the sufficient bouncing motion would require a suitable time period during which resonance may happen. Hopefully, it is understood that SRTM selects both of the frequency of oscillation in the energy level and the damping corresponding to the resonance absorption. In the low energy regime, the selectivity becomes very prominent at the resonance energy \cite{li-2008}, causing a complete suppression of neutron-emission in the SRTM process.

The Maxwellian-averaged thermonuclear reaction rate $<\sigma v >$ at some temperature, T is given by the following integral \cite{boyd-2008}:
\begin{equation}\label{eq01}
<\sigma v> = \sqrt{\frac{8}{\pi\mu (K_BT)^3}}\int\sigma(E)E\exp\left(-\frac{E}{K_BT}\right) dE
\end{equation}
where $E$ is the center-of-mass energy, $v$ is the relative velocity and $\mu$ is the reduced mass of
reactants. At low energies (far below the Coulomb barrier) where the lassical turning point is much larger than the nuclear radius, barrier penetrability can be approximated by $\exp(-2\pi\zeta)$ so that
the charge induced cross section can be decomposed into 
\begin{equation} \label{eq02}
\sigma(E) = \frac{S(E)}{E} \exp(-2\pi\zeta)
\end{equation}
where $S(E)$ is the astrophysical S-factor and $\zeta$ is the Sommerfeld parameter, defined by 
\begin{equation} \label{eq03}
\zeta = \frac{Z_1Z_2e^2}{\hbar v}
\end{equation}
where $Z_1$ and $Z_2$ are the charges of the reacting nuclei in units of elementary charge $e$. Except for narrow resonances, the S-factor S(E) is a smooth function of energy, which is
convenient for extrapolating measured cross sections down to astrophysical energies.
Nuclear fusion reaction in the low energy range ($E\sim$ 1eV to few keV) can be explained successfully by the phenomenon of quantum mechanical tunneling through the mutual Coulomb barrier of interacting nuclide. For two approaching nuclei of charge numbers $Z_1$, $Z_2$, mass numbers $A_1$ and $A_2$, height of the Coulomb barrier is given by

\begin{eqnarray}\label{eq04}
V_{CB}&=&\left(\frac{e^2}{4\pi\epsilon_0}\right)\left(\frac{Z_1Z_2}{R}\right)=
(1.44)\left(\frac{Z_1Z_2}{R_0(A_1^{1/3}+A_2^{1/3})}\right) MeV
\end{eqnarray}
where $R_0$ is the nuclear radius parameter and $R$ is the touching distance between the centers of the interacting nuclei. The cross section for sub-barrier fusion for light nuclei can be calculated using the SRTM assuming a complex nuclear potential (including Coulomb term) of the form
\begin{equation}\label{eq05}
V(r)= V_{N}(r) +V_{C}(r)
\end{equation}
where
\begin{equation}\label{eq06}
V_{N}(r) = -V_{r}\exp\left[  -\left(\frac{r}{\beta_r}\right)^{2}\right] + iV_{i}\exp\left[  -\left(\frac{r}{\beta_i}\right)^{2}\right]
\end{equation}
and
\begin{equation}\label{eq07}
V_C(r) =\left\{ \begin{array}{l}
 1.44 \frac{Z_1Z_2}{2R}\left(3 - \frac{r^2}{R^2}\right), \: \textbf{for}\: r \leq R\\ 
 1.44 \frac{Z_1Z_2}{r}, \: \textbf{for} \:r > R\\ 
\end{array} \right.
\end{equation}
The imaginary component introduced in the nuclear potential facilitates the absorption phenomenon and description of its effect on the associated wave function. The Schor\"{o}dinger equation for the nuclear plus Coulomb potential is given by
\begin{equation}\label{eq08}
\left(-\nabla^2 + \frac{2\mu}{\hbar^2}\left[  V_{N}(r) +V_{C}(r) \right]-k^2
\right)\psi({\bf r})=0
\end{equation}
where $k^2=\frac{2\mu E}{\hbar^2}$ and $\psi({\bf r})$ represents the sum of the nuclear and Coulomb wave function i.e.,
\begin{equation}\label{eq09}
\psi({\bf r}) = \psi_N({\bf r})+\psi_C({\bf r})
\end{equation}
The Coulomb wave function contains the incoming wave while the nuclear wave function represents only the outgoing wave in the asymptotic range. 
For numerical solution of Eq.(\ref{eq08}), it is reduced to the one-dimensional equation in $r$ given by
\begin{equation}\label{eq10}
\left(-\frac{d^2}{dr^2} +\frac{l(l+1)}{r^2}+ \frac{2\mu}{\hbar^2}\left[  V_{N}(r) +V_{C}(r) \right]-k^2
\right)\psi({r})=0
\end{equation}
Thus, when a light nucleus is injected into another light nucleus, the relative motion can be described in terms of the radial wave function $\psi(r)$ connected to the general solution $\psi(r,t)$ of Schr$\ddot{o}$dinger equation for the interacting nuclei as 
\begin{equation}\label{eq11}
\psi(r,t)=\frac{1}{\sqrt{4\pi}r}\psi(r) \exp\left(-i\frac{E}{\hbar}t\right)
\end{equation}
Now, the reaction cross-section in terms of the phase shift, $\delta_0$ introduced by the nuclear potential in the wave function at the low energy limit (where only S-wave contributes) is given by
\begin{equation}\label{eq12}
\sigma=\frac{\pi}{k^2}(1- |\eta|^2)
\end{equation}
where $\eta=e^{2i\delta_0}$ and k is the wave number corresponding to the relative motion.
Since, the chosen nuclear potential is a complex one, the corresponding phase shift $\delta_0$ will also be a complex number and can be expressed as 
\begin{eqnarray}\label{eq13}
\cot(\delta_0)=W_r+i W_i
\end{eqnarray}
where $W_r$ and $W_i$ are two parameters connected to the real and imaginary components of the complex wavenumber corresponding the complex nuclear potential. The wavenumber (K) corresponding to the complex nuclear potential can be expressed as
\begin{equation}\label{eq14}
\left. \begin{array}{lcl}
K &=& \sqrt{\frac{2\mu}{\hbar^2}[V_r(r)+iV_i(r) -E]}\\
  &=& \sqrt{(\frac{2\mu}{\hbar^2})[V_r(r)-E]}[1+i\frac{V_i}{V_r(r) -E}]^{1/2}\\
&=& \sqrt{\frac{2\mu}{\hbar^2}[V_r(r)-E)]} +i \sqrt{\frac{2\mu}{\hbar^2}}\frac{V_i(r)}{2\sqrt{V_r(r)-E}}\\
&=&K_r+iK_i\\
or, \: KR &=& K_rR+iK_iR \: [\textit{Let}, Z = KR; Z_r = K_rR; Z_i = K_iR]\\
\Rightarrow Z&=&Z_r+iZ_i\\
\end{array} \right\}
\end{equation} 
In terms of real and imaginary parts of $Z$ the parameters $W_r$ and $W_i$ are defined as
\begin{equation}\label{eq15}
\left. \begin{array}{lcl}
W_r &=& \chi^2 \left[\left(\frac{R_C}{R}\right)\frac{Z_r\sin(2Z_r)+Z_i\sinh(2Z_i)}{2[\sin^2(Z_r)+\sinh^2(Z_i)]}\right] \\
&&-2\chi^2 \left[\ln\left(\frac{2R}{R_C}\right)+2C+h(kR_C)\right]\\
W_i &=& \chi^2{\it Im}\left[\frac{R_C}{R} (KR) \cot(KR)    \right]\\
&=& \chi^2\left[\frac{R_C}{R}\frac{Z_i\sin(2Z_r)-Z_r\sinh(2Z_i)}{2[\sin^2(Z_r)+\sinh^2(Z_i)]}    \right]
\end{array}  \right\}
\end{equation}
where $R_C = \frac{\hbar^2}{Z_1Z_2\mu e^2}$ is the Coulomb unit of length and $C = 0.577$ is Euler's constant. The function $h(kR_C)$ is connected to the logarithmic derivative of $\Gamma$ function
\begin{equation}\label{eq16}
h(y) = \frac{1}{y^2}\sum_{i=1}^{\infty}\frac{1}{i(i^2+y^{-2})}-C+\ln(y)
\end{equation}
The fusion cross-section can then be expressed as
\begin{equation}\label{eq17}
\left. \begin{array}{lcl}
\sigma&=&\left(\frac{\pi}{k^2}\right)\left(-\frac{4W_i}{(1-W_i)^2+W_r^2}\right)\\
&=&\left(\frac{\pi}{k^2}\right)\left(\frac{1}{\chi^2}\right)\left(-\frac{4\omega_i}{\omega_r^2+ (\omega_i-\frac{1}{\chi^2})^2}\right)\\
\end{array} \right\}
\end{equation}
where the quantity
$\chi^2=\left\{\frac{\exp\left(\frac{2\pi}{kR_C}\right)-1}{2\pi}\right\}$ is related to the Gamow penetration factor. The last factor in Eq. (\ref{eq17}) within curly braces$\{\}$ is called the astrophysical S-factor, which depends on the projectile energy, E and the cross-section $\sigma(E)$. Thus we have
\begin{equation}\label{eq18}
\left. \begin{array}{lcl}
S(E)&=&\left(\frac{k^2}{\pi}\right) 
\left( \chi^2 \right) \sigma(E)\\
&=&\left(-\frac{4\omega_i}{\omega_r^2+ (\omega_i-\frac{1}{\chi^2})^2}\right)
\end{array}  \right\}
\end{equation}
where $\omega=\omega_r+i\omega_i=W/\chi^2=(W_r+iW_i)/\chi^2.$
The wave function inside the nuclear well (ie., in the region r$ < $R) is determined by two depth parameters- the depths of the real and imaginary components of the nuclear potential ($V_{r}$ and $V_{i}$) and corresponding range parameter $\beta_r$ and $\beta_i$ respectively. The Coulomb wave
function outside the nuclear well (r$ > $ R) is determined by two other parameters: the real and the imaginary part of the complex phase shift  $(\delta_{0r})$ and $(\delta_{0i})$. A pair of convenient parameters, $W_r$ and $W_i$, are introduced to make a linkage
between the cross section and the nuclear potential. This facilitates a clear understanding of the resonance and the selectivity in damping.
The continuity of the wave function at the boundary ($r = R$) can be expressed by the matching of the logarithmic derivative of the wave function.
In the above relations, k represents the wavenumber outside the nuclear well and $R_C$ is the Coulomb unit of length.

The cross-section $\sigma$ can also be computed using the following three- and five parameter fitting formulae \cite{li-2008,huba-2013}.
\begin{equation}\label{eq19}
\left. \begin{array}{lcl}
\sigma_3(E_{lab})&=&\left( \frac{\pi}{({\frac{2\mu}{\hbar^2})E_{lab}(\frac{m_2}{m_1+m_2}})}\right)\left(\frac{1}{\chi^2} \right)\\
&&\times \left(\frac{(-4C_3)}{(C_1+C_2E_{lab})^2+(C_3-\frac{1}{\chi^2})^2}\right)\\
\end{array} \right\}
\end{equation}

\begin{equation}\label{eq20}
\left. \begin{array}{lcl}
\sigma_5(E_{lab})&=&\frac{A_5+(A_2/((A_4-A_3E_{lab})^2+1)}{E_{lab}[\exp(A_1/\sqrt{E_{lab}})-1]};\\
E_{lab}&=&(1+\frac{m_1}{m_2})E\\
\end{array} \right\}
\end{equation}

\section{Results and discussion}
In the present SRTM scheme, we  have four prime tuning parameters $V_{r}$ , $V_{i}$, $\beta_{r} $, $ \beta_{i}$ i.e. depths and ranges of real and imaginary components of the chosen nuclear potential. These are adjusted to search the resonance energy. In this case, we fine-tuned these parameters for different nuclear systems to obtain better resonating behavior. Here we also adjust the radius parameter, $R_{0}$ for finding the sharp resonance nature. Here $R_0$ appears in Eq. (\ref{eq07}), Eq. (\ref{eq14}) and Eq. (\ref{eq15}) while defining radius of the nuclear well and height of the Coulomb barrier.  It can be seen that $R_0$ affects the complex phase shift $\delta_0$ defined in Eq. (\ref{eq13}). Which in turn affects the fusion cross-section $\sigma$ given in Eq. (\ref{eq17}) and astrophysical S-factor S(E) presented by Eq. (\ref{eq18}). Thus a change in $R_0$ changes the cross-section as well as the S-factor. Hence, we varied $R_0$ to match our computed data with those obtained by using the three-parameter (Eq. (\ref{eq19})) and five-parameter (Eq. (\ref{eq20})) fitting formula of references \cite{li-2008,huba-2013}. In the case of D-D fusion data-set-0 reflects the resonating the nature of the D-D fusion, while “data-set-1” and “data-set-2” are used to fit the three-parameter and  five-parameter peak values respectively. The fusion cross-sections and astrophysical S-factor are calculated using Eqs. (\ref{eq16}) \& (\ref{eq17}) respectively.

\begin{table}[htbp]
\caption[]{Numerical values of the adjustable potential parameters involved in Eqs. (\ref{eq04}) \& (\ref{eq06}).}\small\smallskip
\tabcolsep=4.6pt
\begin{tabular}{@{}ccccccc@{}}
\hline
&&&&&\\[-10pt]
\textbf{Reactions}&\textbf{$V_{r}$}   &\textbf{$V_{i}$}&\textbf{$\beta_{r}$}   &\textbf{$\beta_{i}$}&\textbf{$R_{0}$}&Reference \\
\textbf{}    &\textbf{(MeV)}&\textbf{(MeV)}&\textbf{(fm)}&\textbf{(fm)}  &\textbf{(fm)}&label \\
\hline
&&&&&\\
[-10pt]D(D, n)$^{3} $He &-36.135 &-0.1330 &1.25&1.50& 2.77795&data-set-0 \\
&&&\\
[-10pt] &-36.135 &-0.2528 &0.42&1.50& 2.30795&data-set-1  \\
&&&\\
[-10pt]&-36.135 &-0.2625 &0.50&1.50& 2.29795&data-set-2 \\
&&&\\
[-10pt]$^{11}$B(p, $\alpha)^8$Be &-35.700&-0.0129&0.50&1.50& 1.7979&data-set-0   \\
&&&\\
[-10pt]&-35.600&-0.05825&0.50&1.50& 1.88620&data-set-1  \\
\hline
\end{tabular}
\label{t01}
\end{table}

\begin{figure}
	\centering
	\fbox{\includegraphics[width=0.45\linewidth, height=0.35\linewidth]{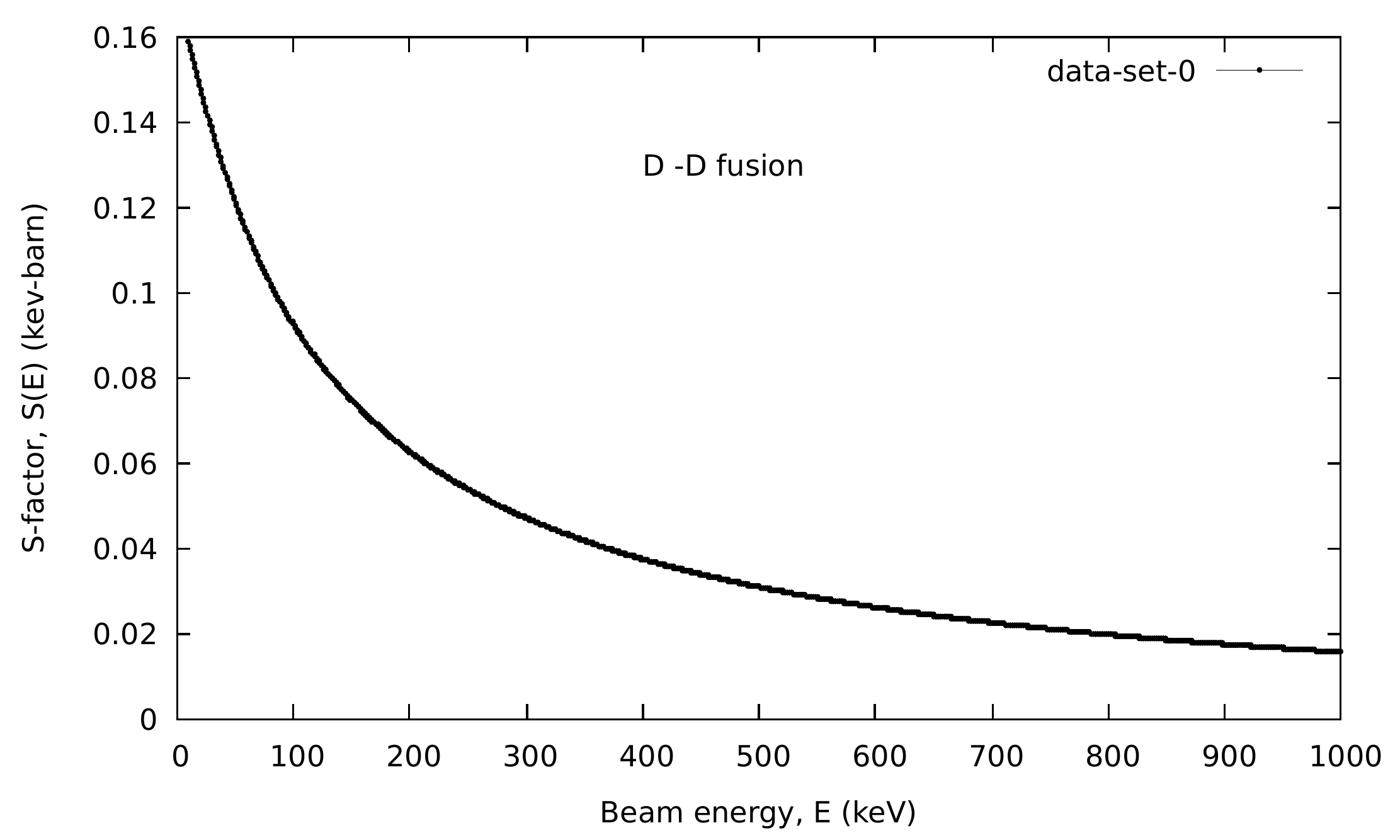}}
	\fbox{\includegraphics[width=0.45\linewidth, height=0.35\linewidth]{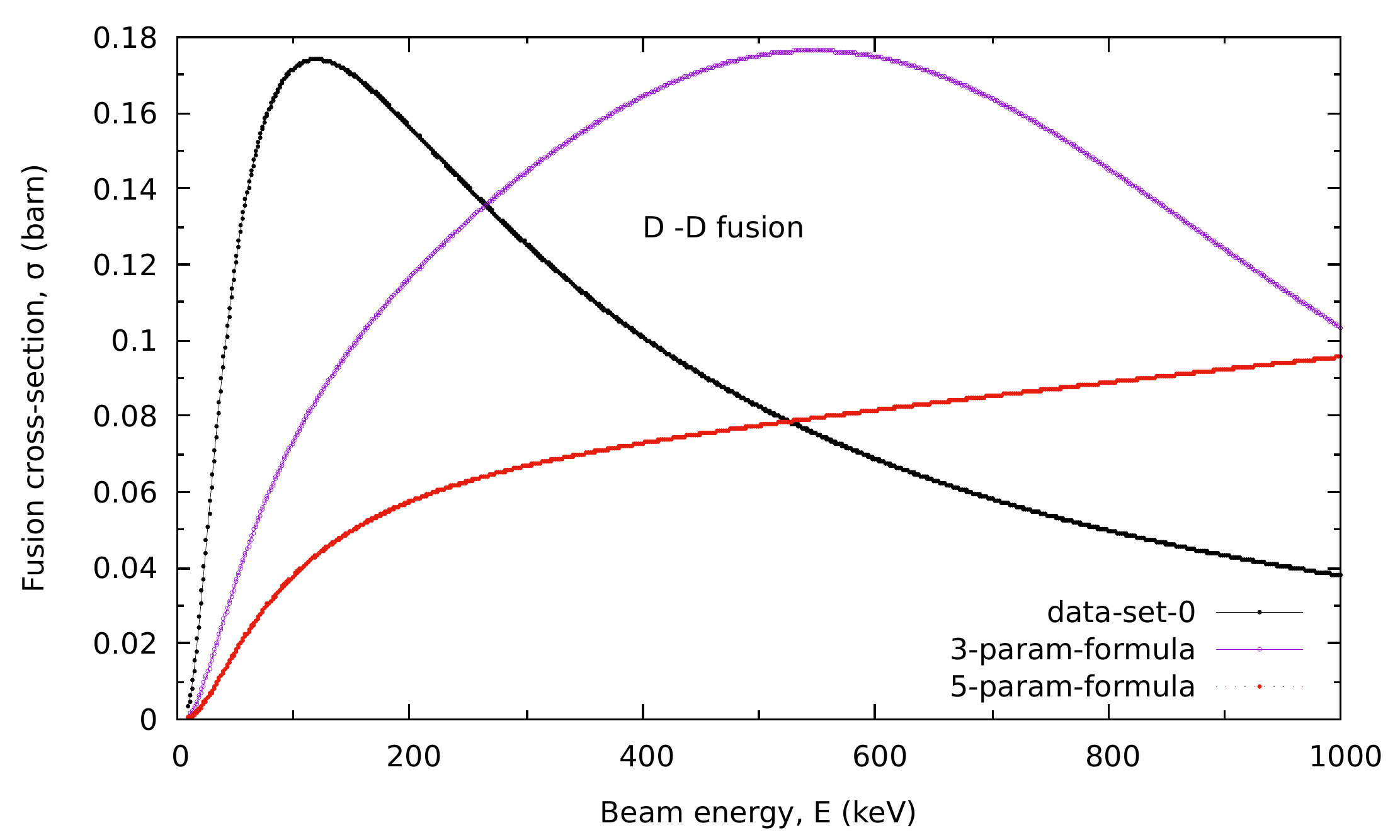}}
	\caption{(Left) S-factor S(E) in \textbf{KeV-barn} calculated for D-D fusion reaction using Eq. (\ref{eq18}), (Right) Comparison of Fusion cross section, $\sigma$ in \textbf{barn} calculated by the use of Eqs. (\ref{eq17}), (\ref{eq19}) \& (\ref{eq20}).}
	\label{fig01}
\end{figure}

\begin{figure}
	\centering
	\fbox{\includegraphics[width=0.46\linewidth, height=0.35\linewidth]{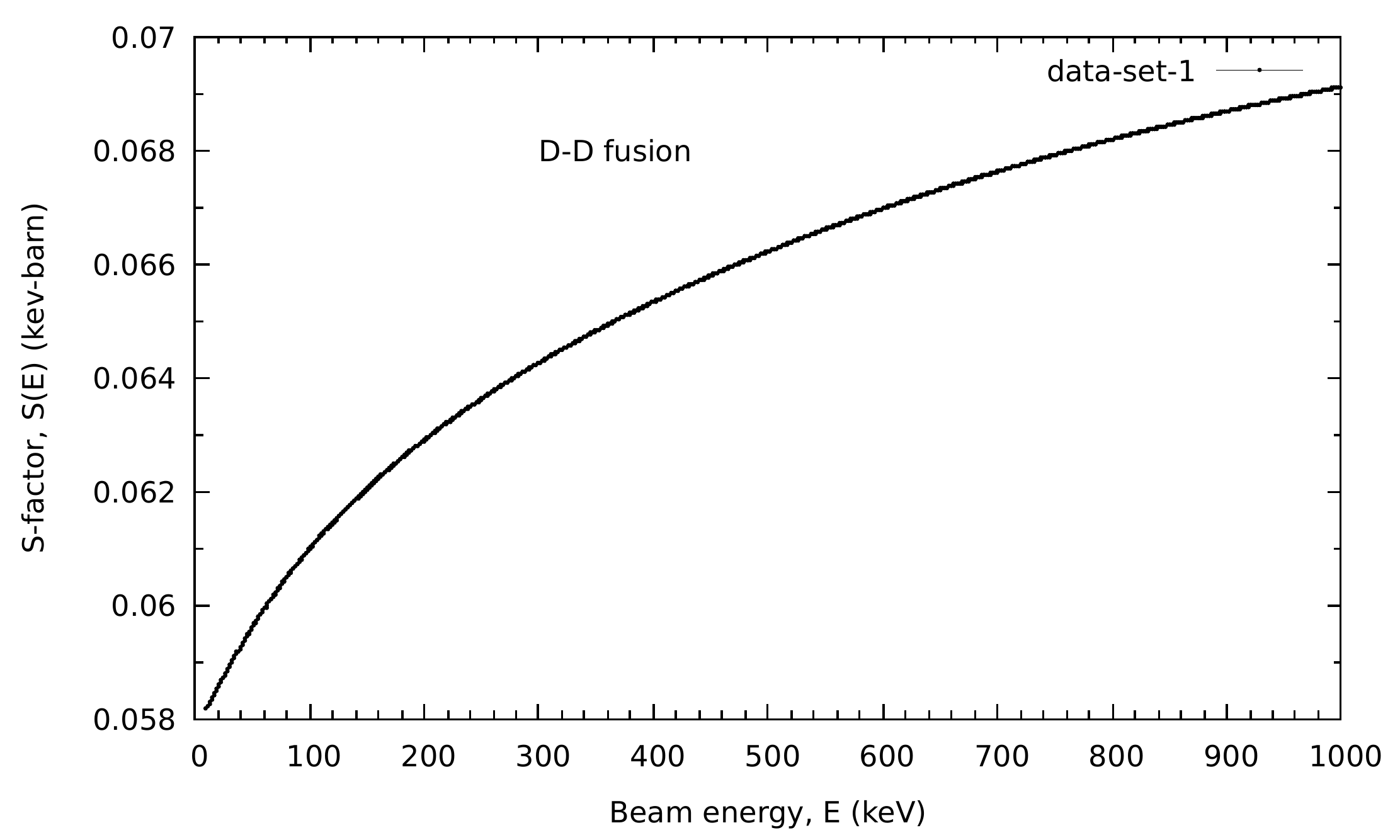}}
	\fbox{\includegraphics[width=0.46\linewidth, height=0.35\linewidth]{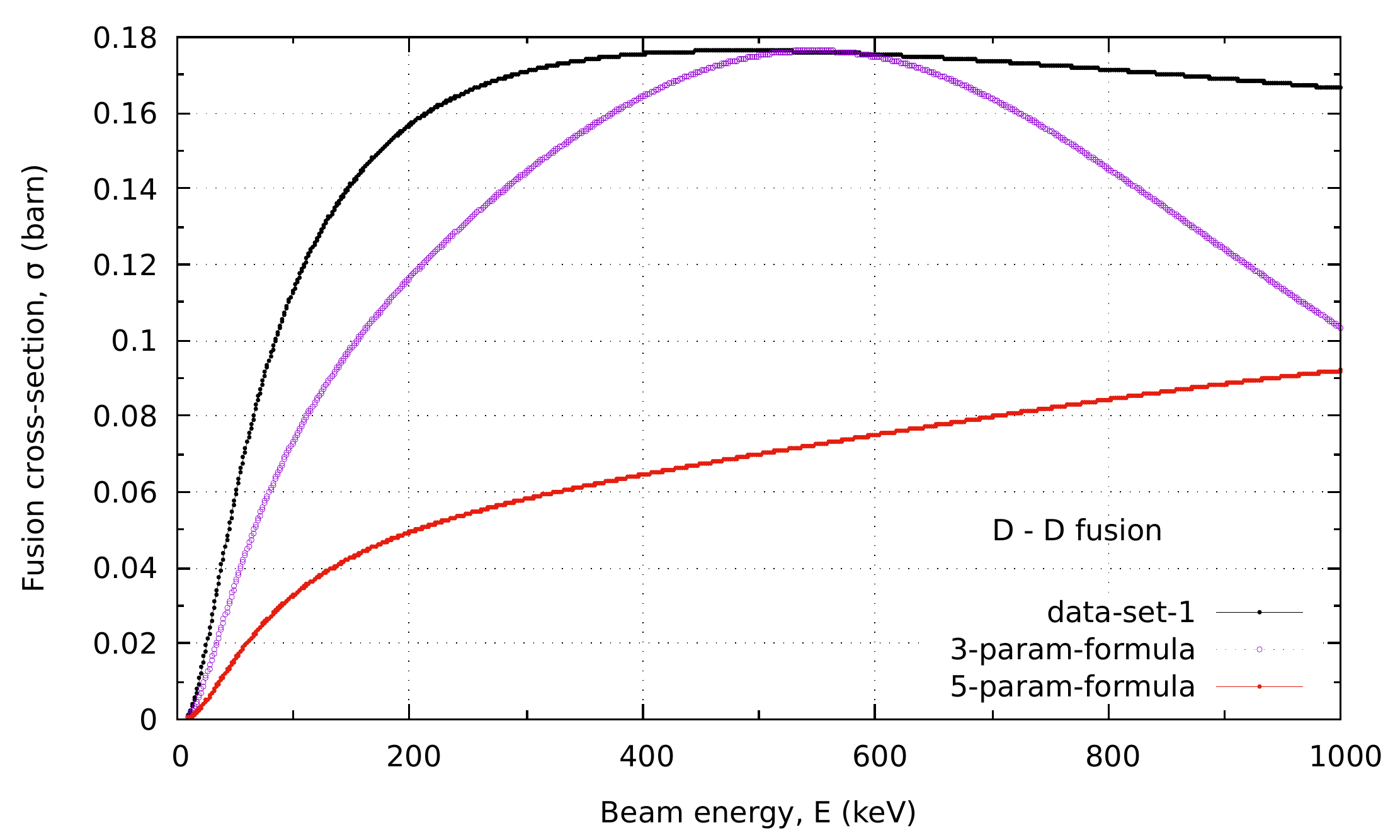}}
	\caption{(Left) S-factor S(E) in \textbf{KeV-barn} calculated for D-D fusion reaction using Eq. (\ref{eq18}), (Right) Comparison of Fusion cross section, $\sigma$ in \textbf{barn} calculated by the use of Eqs. (\ref{eq17}), (\ref{eq19}) \& (\ref{eq20}).}
	\label{fig02}
\end{figure}

\begin{figure}
	\centering
	\fbox{\includegraphics[width=0.45\linewidth, height=0.35\linewidth]{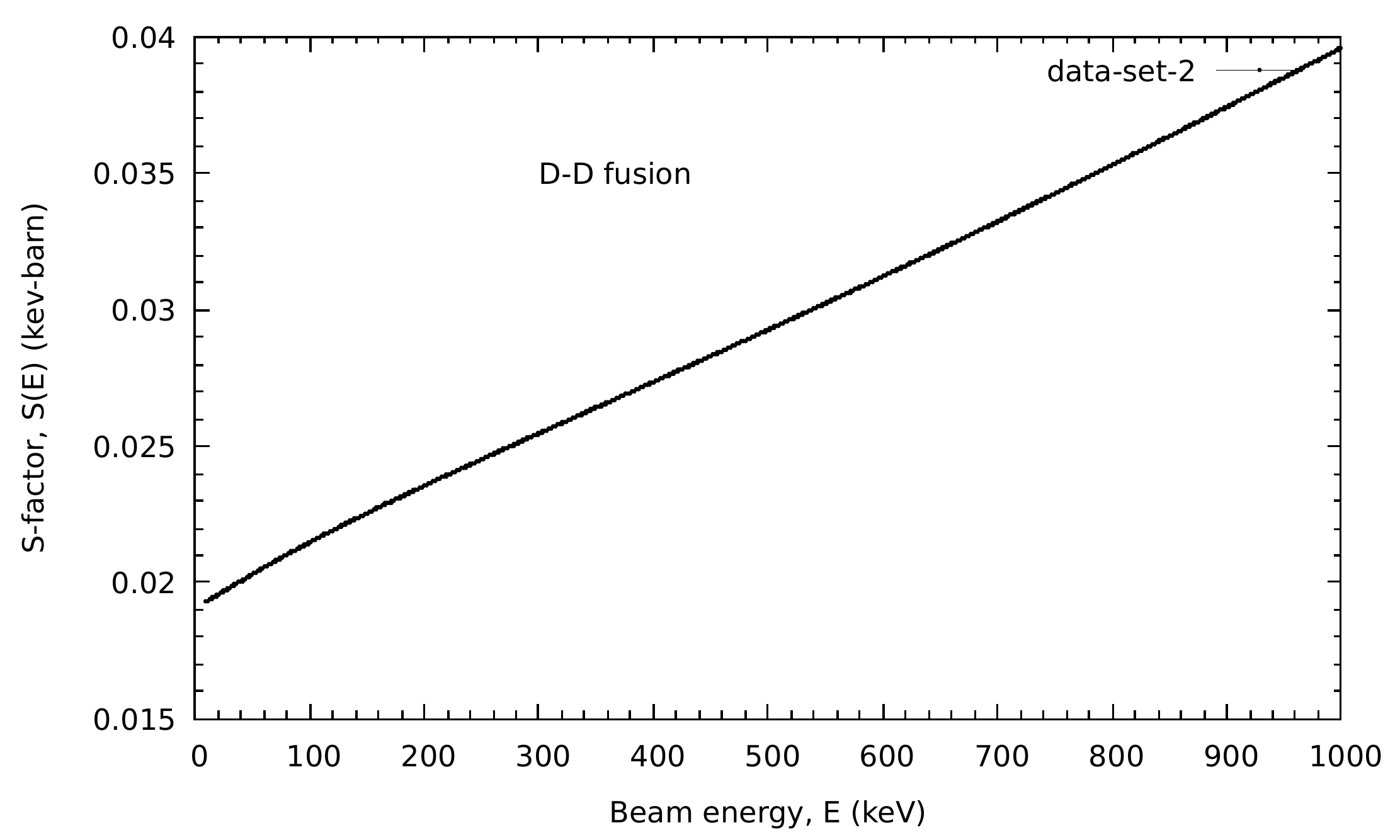}}
	\fbox{\includegraphics[width=0.45\linewidth, height=0.35\linewidth]{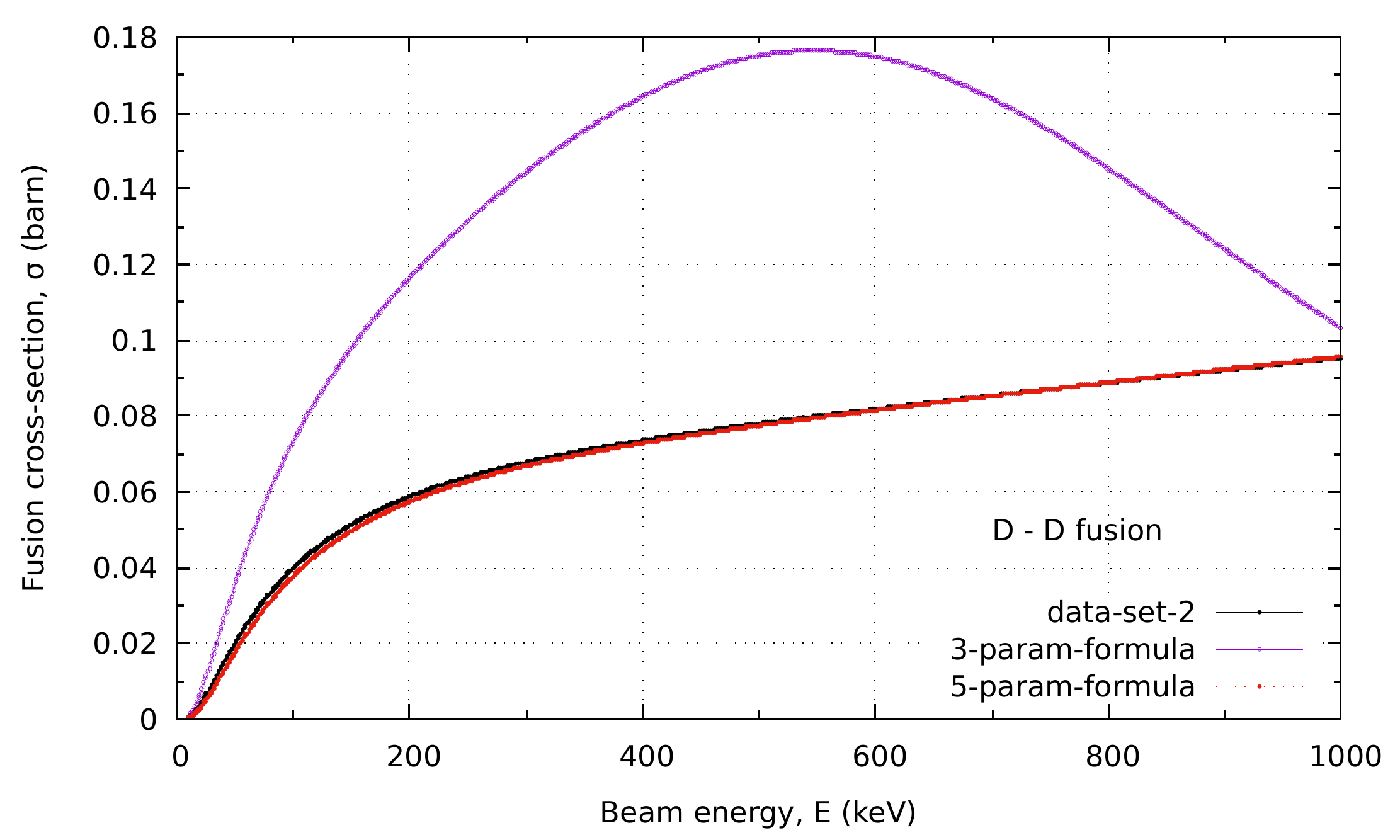}}
	\caption{(Left) S-factor S(E) in \textbf{KeV-barn} calculated for D-D fusion reaction using Eq. (\ref{eq18}), (Right) Comparison of Fusion cross section, $\sigma$ in \textbf{barn} calculated by the use of Eqs. (\ref{eq17}), (\ref{eq19}) \& (\ref{eq20}).}
	\label{fig03}
\end{figure}

\begin{figure}
	\center
	\fbox{\includegraphics[width=.45\linewidth, height=0.35\linewidth]{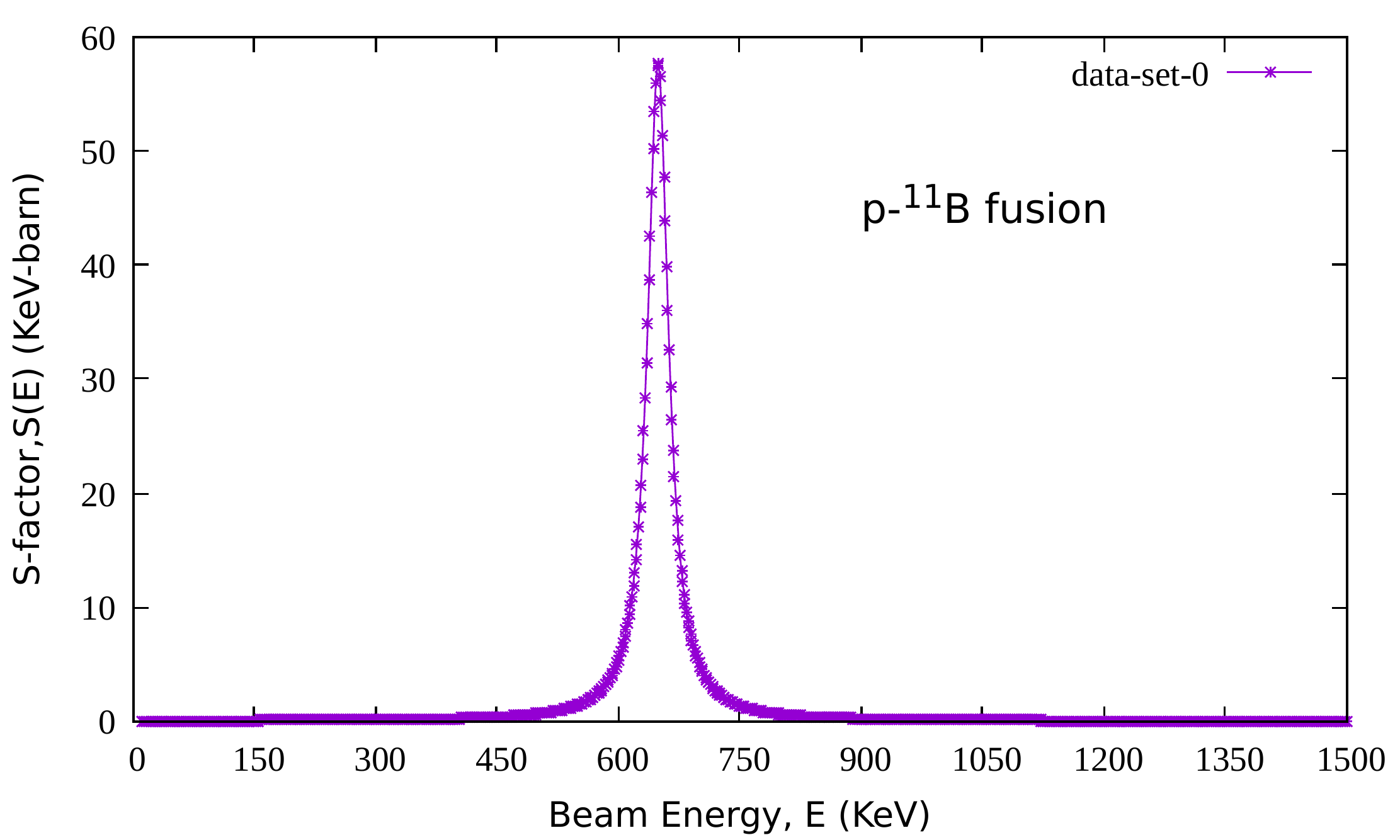}}
	\fbox{\includegraphics[width=.45\linewidth, height=0.35\linewidth]{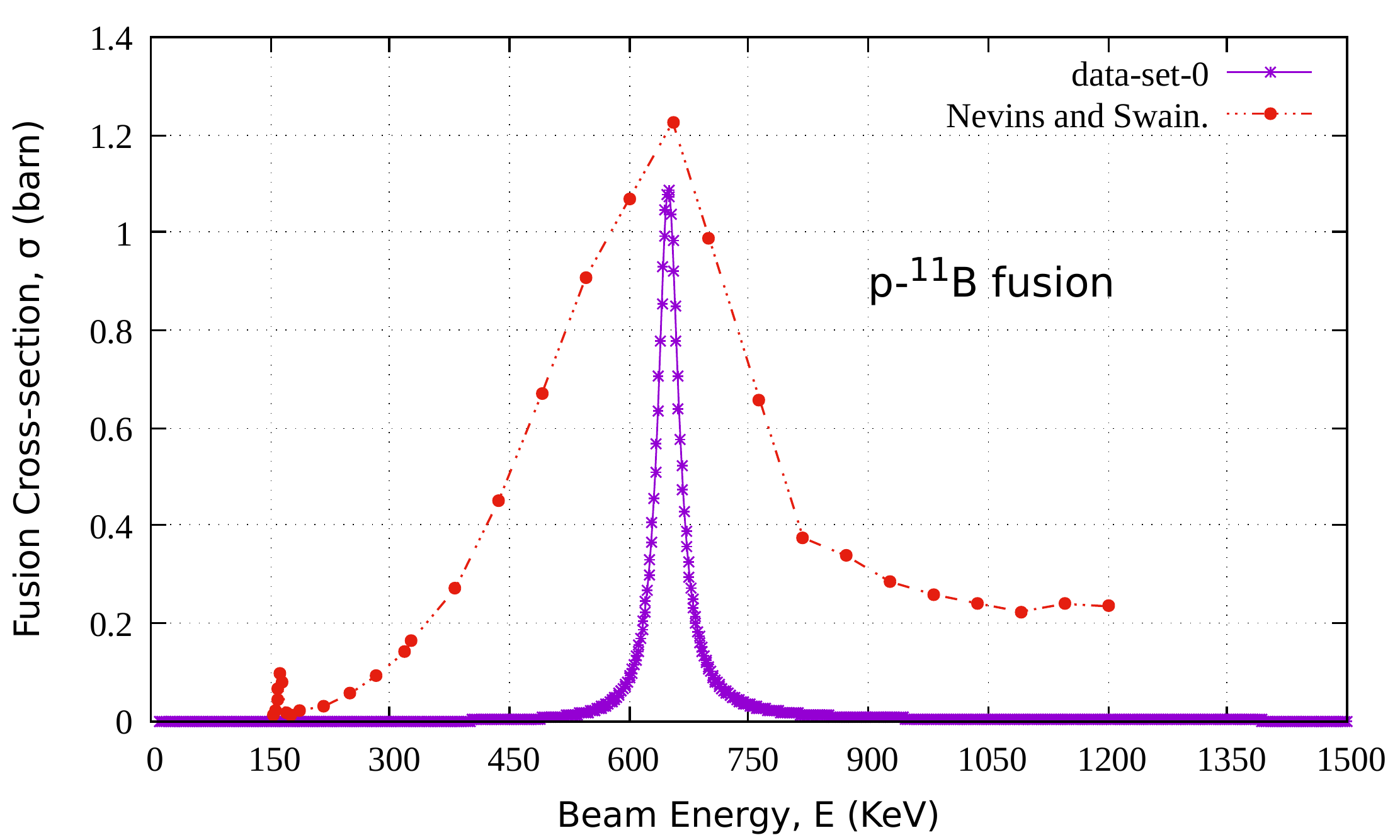}}
	\caption{(Left) S-factor S(E) in \textbf{KeV-barn} calculated for p-$^{11}$B fusion reaction using Eq. (\ref{eq18}), (Right)  Comparison of Fusion cross section, $ \sigma $ in \textbf{barn} obtained by Eq.(\ref{eq17}) (solid line) with those reported by Nevins and Swain 2000 \cite{nevins-2000}, Becker et al 1987 \cite{becker-1987} (dashed line).}
	\label{fig04}
\end{figure}

\begin{figure}
	\center
	\fbox{\includegraphics[width=.45\linewidth, height=0.35\linewidth]{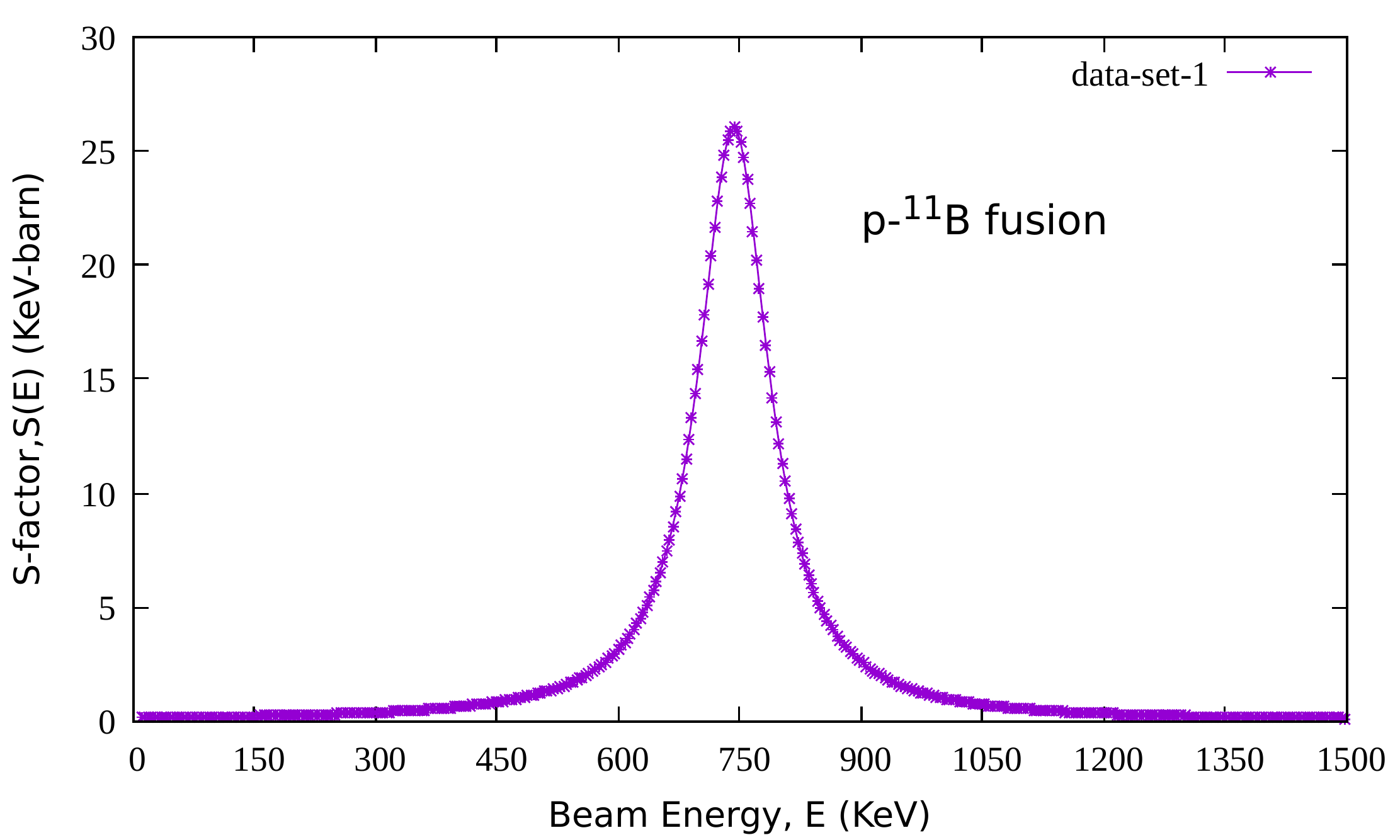}}
	\fbox{\includegraphics[width=.45\linewidth, height=0.35\linewidth]{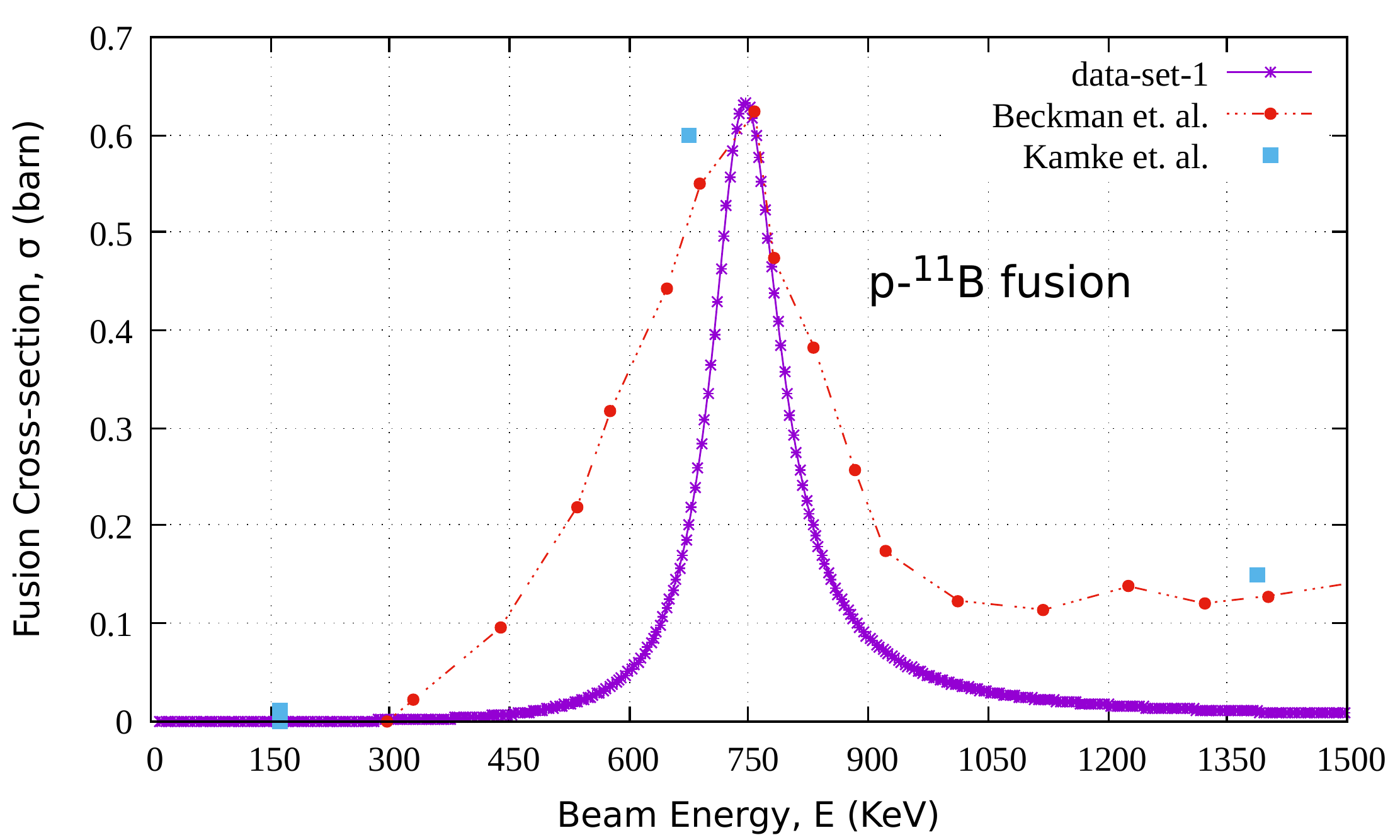}}
	\caption{(Left) S-factor S(E) in \textbf{KeV-barn} calculated for p-$^{11}$B fusion reaction using Eq. (\ref{eq18}), (Right)  Comparison of Fusion cross section, $ \sigma $ in \textbf{barn} obtained by Eq. (\ref{eq17}) (solid line) with those reported by Beckman et al 1953 \cite{beckman-1953} (dashed line) and Kamke and Krug 1967 \cite{kamke-1967}(filled square).} 
    \label{fig05}
\end{figure}

The fusion cross-section, $\sigma$ and the astrophysical S- function are computed by the use of Eqs. (\ref{eq17}) \&  (\ref{eq18}) respectively at energies below the height of the barrier. Plot of the astrophysical S-factor and fusion cross-sections against energy of the projectile are shown in Figures \ref{fig01}, \ref{fig02}, \ref{fig03}, \ref{fig04} \& \ref{fig05} respectively for five different sets of parameters listed in Table \ref{t01}. 

The first three-rows of the parameter values in Table  \ref{t01} are used for the $D-D$ fusion reaction while the fourth row is used for the p-$^{11}$B reaction. The second and third sets of parameters (namely, data-set-1 and data-set-2) for the $D-D$ fusion reaction are chosen to obtain data that are comparable to those obtained by the use of 3-parameter and 5-parameter formulae (see Eqs. (\ref{eq19}) \&  (\ref{eq20})) found in the literature \cite{li-2008,huba-2013}. Results of the $D-D$ fusion obtained with the set of parameters presented in the second row of Table \ref{t01} agrees nicely with the result of the 3-parameter formula (see Eq. (\ref{eq19})) as depicted in the right panel of Figure \ref{fig02}. Similarly, the same obtained with the set of parameters presented in the third row of Table \ref{t01} agrees excellently with the result of the 5-parameter formula (see Eq. (\ref{eq20})) as depicted in the right panel of Figure \ref{fig03}. However, those two sets of data do not exhibit resonance behavior clearly in the cross-section versus energy graph. Hence, we fine-tuned the adjustable parameters in Eqs.(\ref{eq04}) \& (\ref{eq06}) to obtain the set of data (labelled as \enquote{data-set-0}) presented in the first row of Table \ref{t01}, which indicate a prominent resonance absorption near 120 keV with an approximate fusion cross-section of 0.174 barn. We used two sets of parameters (namely, data-set-0 \& data-set-1) for p-$^{11}$B cross-section and s-function calculation which are respectively presented in the fourth and fifth rows of Table \ref{t01}.
Cross-section versus energy graph for p-$^{11}$B obtained with the set of parameters presented in the fourth row of Table \ref{t01} exhibits a sharp peak at around 650 keV indicating a clear resonance with a cross-section of about 1.1 barn which is in good agreement with the results reported by Nevins and Swain 2000 \cite{nevins-2000}, Becker 1987 \cite{becker-1987} in which they obtained a peak cross-section of about 1.2 barn near 654 keV as shown in Figure \ref{fig04}. In Figure \ref{fig05} plot of cross-section versus energy for p-$^{11}$B is obtained with the set of parameters presented in the fifth row of Table \ref{t01}. The plot exhibits a sharp resonance peak near 760 keV with a peak fusion cross-section value of about 0.62 barn which in excellent agreement with the results reported by Beckman et al 1953 \cite{beckman-1953} who reported a peak cross-section of about 0.623 barn near 757 keV. The astrophysical S-function depicted in the left panels of Figures \ref{fig04} and \ref{fig05} indicate a clear resonance behavior for both the chosen data sets. The quantum-mechanical computation agrees well with the experimental findings. The experimental data for both of the D-D and p-$^{11}$B fusion are well reproduced by our chosen interaction. For D-D fusion our potential model indicates clear selective resonance around 120 keV (\textbf{119.98 keV, 0.174 barn}) for data-set-0 (Figure \ref{fig01}). And for p-$^{11}$B, a more prominent selective resonance around 650 KeV (\textbf{647.22 keV, 1.09 barn}) for the data-set-0 (Figure \ref{fig04}) and around 760 keV (\textbf{760.024 keV, 0.635 barn}) for data-set-1 (Figure \ref{fig05}) respectively. Yet calculations for fusion cross-section involving heavy and medium mass nucleus-nucleus systems, need a different treatment \cite{atta-2014}.

\section{Summary and conclusion}
The compound nuclear model fails to describe the process of fusion reaction of light nuclei at very low energies, since the fusing nuclei may still
remember the phase factor of the wave function describing the system.
In the compound nuclear model, the reaction is assumed to proceed in two
steps: first fusing to form the compound nucleus followed by its decay.
Here we consider the selective resonant tunneling model according to which the tunneling probability itself depends upon the decay lifetime, hence it is a single-step process. The agreement with the experimental data for the deep sub-barrier fusion of light nuclei also suggests that the tunneling proceeds in a single step.

The present study successfully reproduces the expected result as evident from the sharp resonance peak in the calculated cross-sections (see Figures \ref{fig01}, \ref{fig04} \& \ref{fig05}). This method (SRTM) also indicates the soundness in explaining light nuclear fusion reactions by the present tunneling model using a complex nuclear potential function. As the fusion cross-section obtained from thermonuclear reaction rate calculation is found to be too small to measure in a laboratory, the present method could be more fruitful to achieve the desired goal.

\section*{Acknowledgements} Authors acknowledge a fruitful discussion with D. N. Basu, V. Singh, and D. Atta of VECC Kolkata. And also Aliah University for providing computational facilities.


\end{document}